\documentclass{amsart}

\usepackage[round]{natbib}
\usepackage{amsmath,amsfonts,amssymb,amsthm}
\usepackage{graphicx}

\newtheorem{defn}{Definition}

\newcommand{\EE}{\mathbb{E}}

\newcommand{\guppy}{\textsf{guppy}}
\newcommand{\pplacer}{\textsf{pplacer}}

\newcommand{\Var}{\operatorname{Var}}
\newcommand{\Cov}{\operatorname{Cov}}

\newcommand{\Sij}{S_{i,j}}
\newcommand{\Sji}{S_{j,i}}
\newcommand{\Oij}{O_{i,j}}
\newcommand{\Oji}{O_{j,i}}
\newcommand{\Xr}{X^r}
\newcommand{\Xu}{X^u}
\newcommand{\Yr}{Y^r}
\newcommand{\Yu}{Y^u}
\newcommand{\qbarD}{{}^q \! \bar{D}}

\newcommand{\eat}[1]{}

\renewcommand{\section}[1]{%
\bigskip
\begin{Large}
\noindent\normalfont\scshape #1
\medskip
\end{Large}
}
\renewcommand{\subsection}[1]{%
\bigskip
\begin{large}
\noindent\normalfont\itshape #1
\end{large}
}
\renewcommand{\subsubsection}[1]{%
\vspace{2ex}
\noindent
\textit{#1.}---}
\newcommand{\arxiv}[1]{#1}
\newcommand{\notarxiv}[1]{}



\newcommand{\FIGdefs}{\
\begin{figure}[ht]
\begin{center}
  \includegraphics[width=7cm]{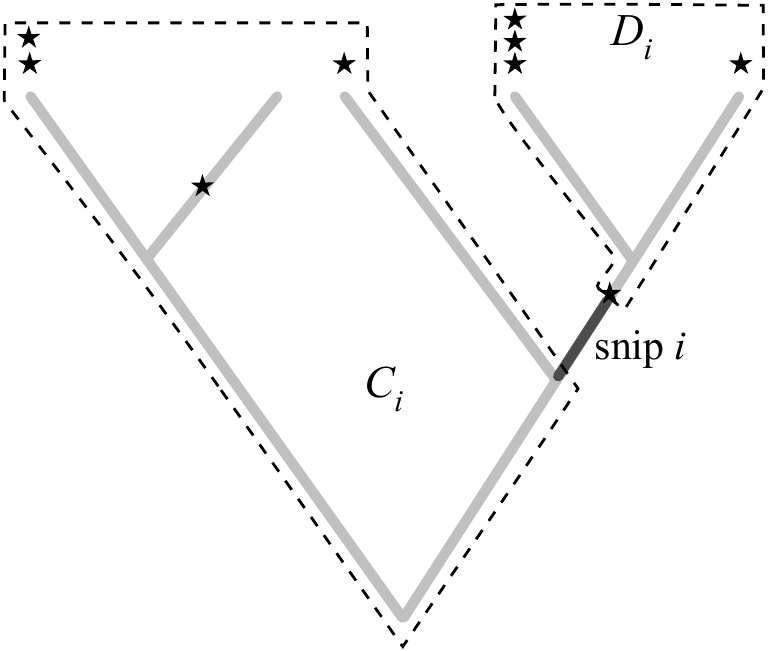}
\end{center}
\caption{\
A hypothetical phylogenetic tree illustrating key concepts in the formulation of the rarefaction of phylogenetic diversity.
The tree is populated with marks (indicated by stars) which represent observations of particular points on the tree in a sample.
Marks might commonly be placed only at the leaves (tips) of the tree but allowing marks to occur anywhere provides for more flexible applications.
Multiple marks indicate multiple observations: for example, several individuals of a species.
The tree can then be broken up into \emph{snips}, which are the edge segments between marks and/or internal nodes.
For each snip $i$, there are two sets of marks, $C_i$ and $D_i$, which name the set of marks that are on the proximal (towards the root) side of $i$ versus those on the distal (towards the leaves) side of $i$.
}
\label{FIGdefs}
\end{figure}
}

\newcommand{\FIGmean}{\
\begin{figure}[ht]
\begin{center}
  \includegraphics[width=7cm]{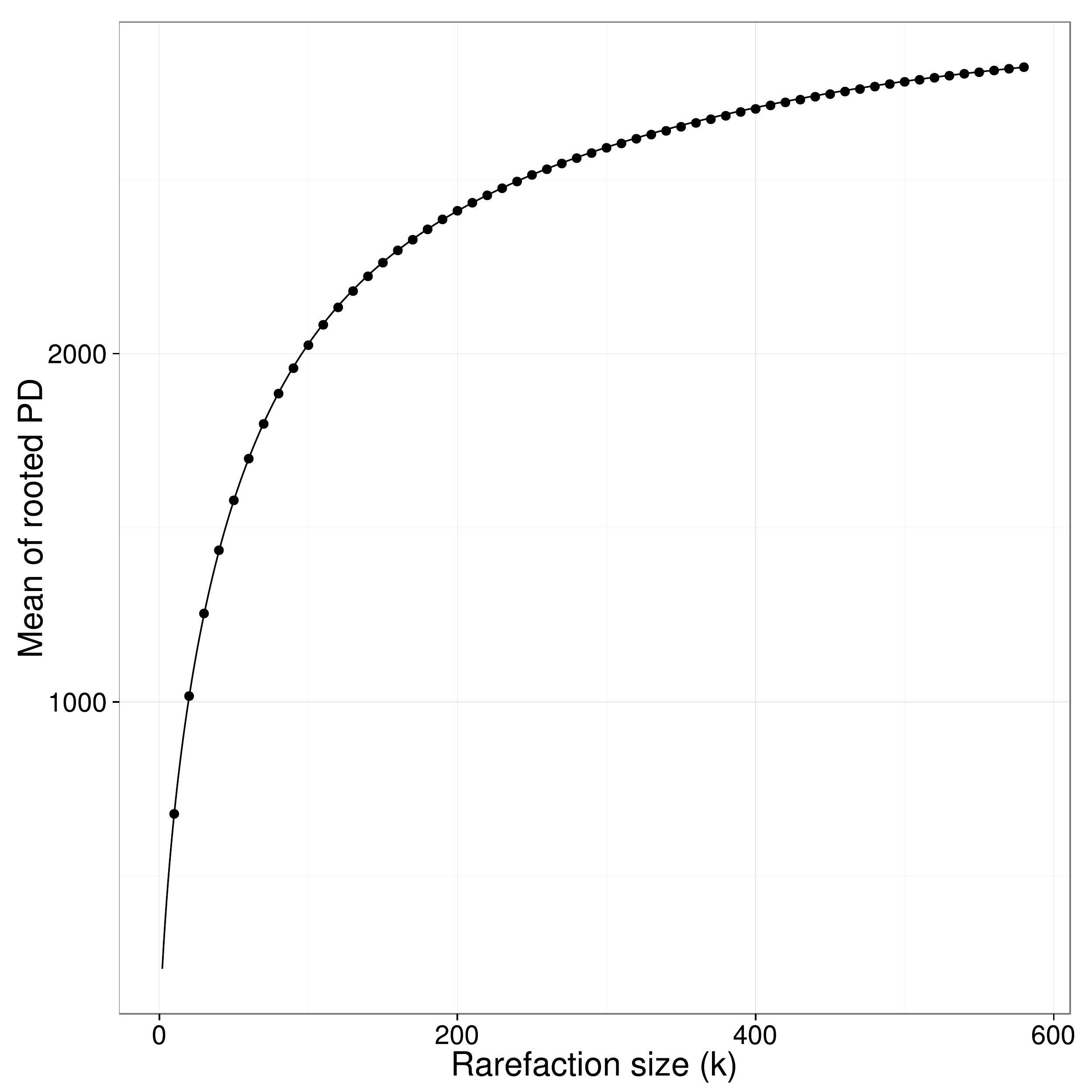}
\end{center}
\caption{\
  Comparison of analytical value (curve) with Monte Carlo calculation with 2,000 samples (points) for the mean of rooted PD under rarefaction.
}
\label{FIGmean}
\end{figure}
}

\newcommand{\FIGvariance}{\
\begin{figure}[ht]
\begin{center}
  \includegraphics[width=7cm]{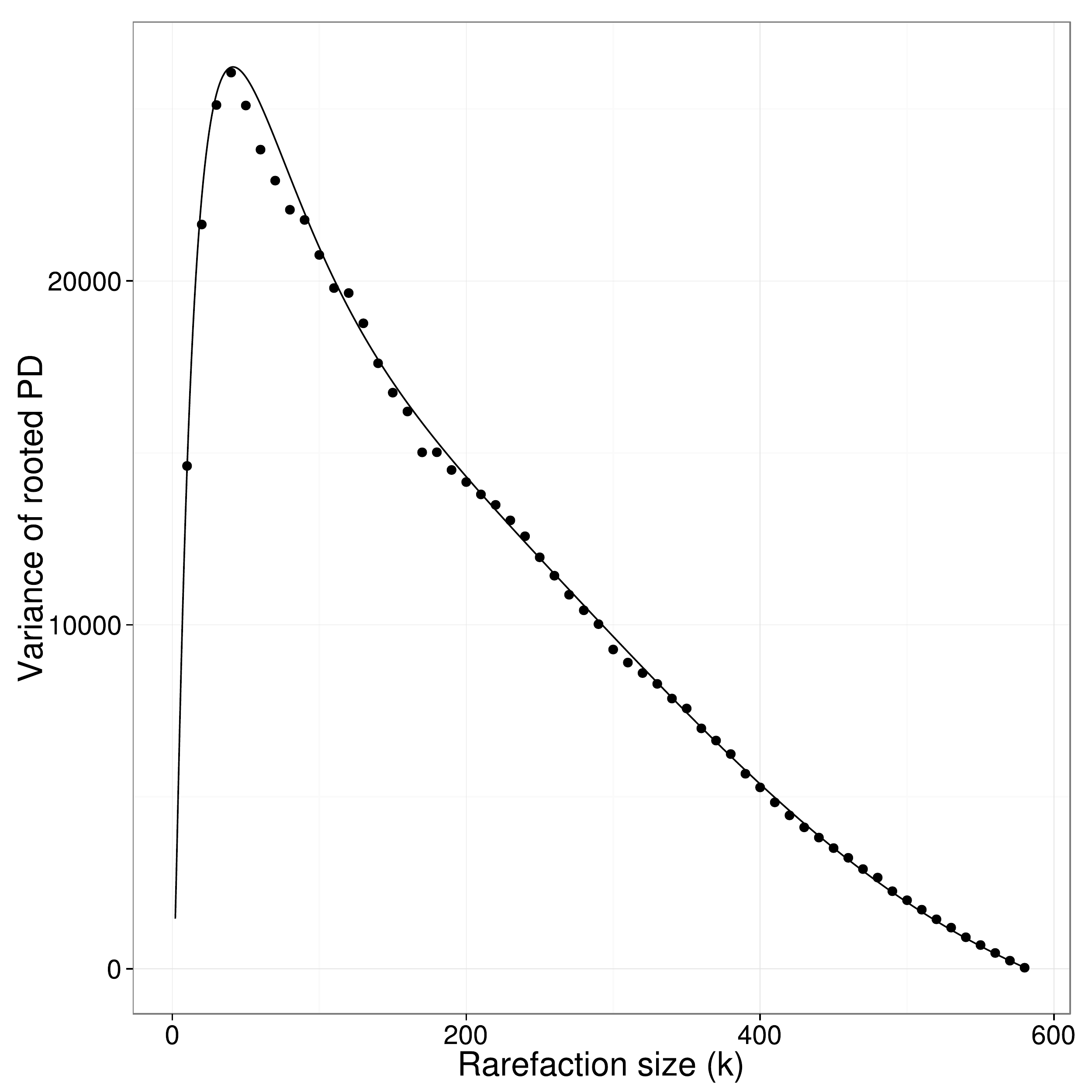}
\end{center}
\caption{\
  Comparison of analytical value (curve) with Monte Carlo calculation with 2,000 samples (points) for the variance of rooted PD under rarefaction.
}
\label{FIGvariance}
\end{figure}
}

\newcommand{\FIGmammalmap}{\
\begin{figure}[ht]
\begin{center}
  \includegraphics[width=12cm]{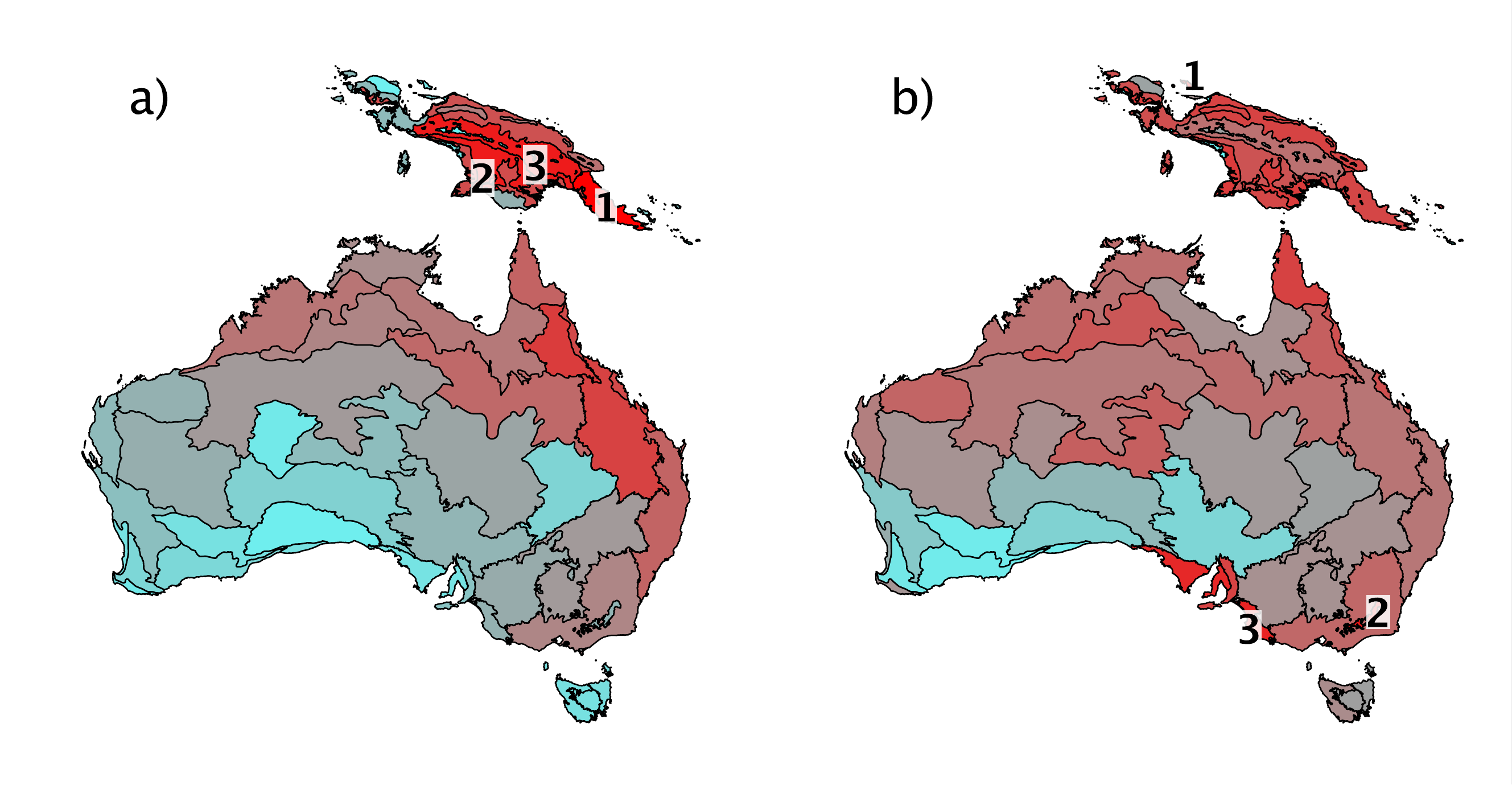}
\end{center}
\caption{\
  Phylogenetic diversity of mammal faunas for terrestrial ecoregions on the Australian continental shelf.
  Phylogenetic diversity is calculated for (a) all species present and (b) as an expected value after rarefaction to 25 species.
  Ecoregions are coloured light blue for low values to dark red for high values.
  The three highest ranked ecoregions in each case are indicated by number.
}
\label{FIGmammalmap}
\end{figure}
}

\newcommand{\FIGrarefactNugent}{\
\begin{figure}[ht]
\begin{center}
  \includegraphics[width=11cm]{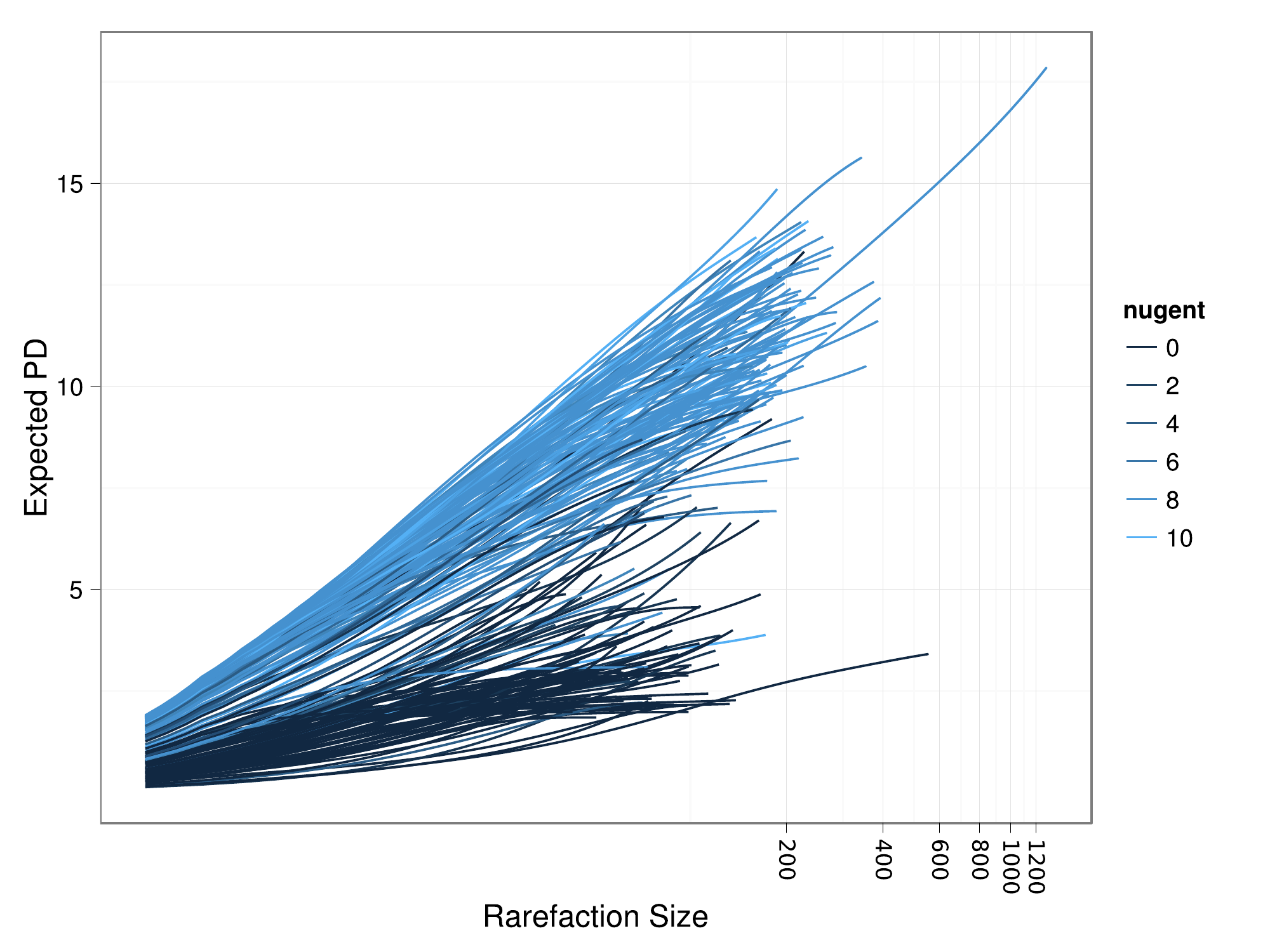}
\end{center}
\caption{\
Rarefaction curve of samples from \citep{srinivasan2012bacterial}.
The Nugent score is a diagnostic score for bacterial vaginosis, with 0 being ``normal'' and 10 being classified as BV.
}
\label{FIGrarefactNugent}
\end{figure}
}

\begin{document}

\begin{titlepage}
  \begin{center}
    {\LARGE  The mean and variance of phylogenetic diversity under rarefaction \par}%
    \vspace{3em}
    {\large
      David A. Nipperess \par Department of Biological Sciences \par Faculty of Science \par Macquarie University, NSW, 2109 \par Australia \par + 61 2 9850 6950 \par david.nipperess@mq.edu.au

     \vspace{1.5em}

      Frederick A. Matsen IV \par Computational Biology Program \par Fred Hutchinson Cancer Research Center \par 1100 Fairview Ave. N. \par Mail stop: M1-B514 \par Seattle, WA 98109-1024 \par +1 206 667 7318

     \vspace{1.5em}

     Corresponding author: \par Frederick A. Matsen IV \par matsen@fhcrc.org \par Fax: +1 206 667 2437

     \vspace{1.5em}
     Running title: RAREFACTION OF PHYLOGENETIC DIVERSITY

     \vspace{1.5em}
    Word count: 4425
    }
  \end{center}\par
\end{titlepage}

\section{Summary}


\begin{enumerate}
  \item Phylogenetic diversity (PD) depends on sampling depth, which complicates the comparison of PD between samples of different depth. One approach to dealing with differing sample depth for a given diversity statistic is to rarefy, which means to take a random subset of a given size of the original sample. Exact analytical formulae for the mean and variance of species richness under rarefaction have existed for some time but no such solution exists for PD.
  \item We have derived exact formulae for the mean and variance of PD under rarefaction. We confirm that these formulae are correct by comparing exact solution mean and variance to that calculated by repeated random (Monte Carlo) subsampling of a dataset of stem counts of woody shrubs of Toohey Forest, Queensland, Australia. We also demonstrate the application of the method using two examples: identifying hotspots of mammalian diversity in Australasian ecoregions, and characterising the human vaginal microbiome.
  \item There is a very high degree of correspondence between the analytical and random subsampling methods for calculating mean and variance of PD under rarefaction, although the Monte Carlo method requires a large number of random draws to converge on the exact solution for the variance.
  \item Rarefaction of mammalian PD of ecoregions in Australasia to a common standard of 25 species reveals very different rank orderings of ecoregions, indicating quite different hotspots of diversity than those obtained for unrarefied PD. The application of these methods to the vaginal microbiome shows that a classical score used to quantify bacterial vaginosis is correlated with the shape of the rarefaction curve.
  \item The analytical formulae for the mean and variance of PD under rarefaction are both exact and more efficient than repeated subsampling. Rarefaction of PD allows for many applications where comparisons of samples of different depth is required.
\end{enumerate}

\section{Keywords}

alpha diversity; phylogenetic diversity; rarefaction; sampling depth

\newpage

\section{Introduction}

Phylogenetic Diversity (PD), the total branch length of a phylogenetic tree, has been extensively used as a measure of biodiversity.
Originally conceived of as a method for prioritising regions for conservation \citep{Faith:1992vm}, PD has seen wider use in other applications such as biogeography \citep{Davies:2011hz}, macroecology \citep{Meynard:2011fl} and microbial ecology \citep{lozupone2008species,turnbaugh2008core,pmid22071345,Yu:2012jg,phillips2012microbiome}.
This increasing breadth of application can be attributed to a number of desirable properties including: 1) explicitly addressing the non-equivalence of species in their contribution to overall diversity, 2) acting as a surrogate for other aspects of diversity such as functional diversity \nocite{Cadotte:2009wb,faith1996conservation} (Cadotte et al., 2009, but see also Faith, 1996), 3) incorporating information on the evolutionary history of communities and biotas and 4) being robust to problems of species delineation because the relationships between populations and even individuals can be represented by relative branch lengths without the need to establish absolute species identity.
Further, the original simple formulation of \citet{Faith:1992vm} has been built on to produce a broader ``PD calculus'' measuring such aspects of diversity as phylogenetic endemism \citep{Faith:2004wc,Rosauer:2009ci}, evenness \citep{hill1973diversity,allen2009new} and resemblance \citep{ferrier2007using,lozupone2008species,Faith:2009tj,Nipperess:2010kp}.
For the purposes of this paper, when referring to ``phylogenetic diversity" and ``PD", we refer explicitly to the definition of \citet{Faith:1992vm}, where diversity is measured as the sum of branch lengths of a phylogenetic tree.

Phylogenetic diversity increases with increasing sampling effort just like many other measures of biodiversity.
Thus, the comparison of the phylogenetic diversity of communities is not straightforward when sample sizes differ, as is common with real data sets.
Unless data are standardised in some sense to account for differences in sample size or effort, the relative diversity of communities can be profoundly misinterpreted \citep{Gotelli:2001uo}.

The established solution to the problem of interpreting diversity estimates with samples of varying size is rarefaction.
The rarefaction of a given sample of size $n$ to a level $k$ is simply the uniform random choice of $k$ of the $n$ observations (typically without replacement).
The observations are typically of either individual organisms or collections of organisms, giving either individual-based or sample-based rarefaction curves \citep{Gotelli:2001uo}.
To consider a given measure of diversity under rarefaction, the measure of diversity is simply applied to the rarefied sample.
Researchers are typically interested in the expectation and variance of a measure of diversity under rarefaction.

Rarefaction curves can be used to understand the depth of sampling of a community compared to its total diversity.
Additionally, rarefaction curves capture information about evenness \citep{olszewski2004unified} and beta-diversity \citep{crist2006additive}, depending on whether observations are of individuals or collections.
Rarefaction curves have been computed for phylogenetic diversity \citep{lozupone2008species,turnbaugh2008core,pmid22071345,Yu:2012jg}.
In each of these cases, rarefaction was not by counts of individual organisms or collections of such, but was instead based on counts of unique sequences or Operational Taxonomic Units.
Rarefaction by such units, including taxonomic species, makes sense in the context of phylogenetic diversity where it might not with other measures of biodiversity.
In effect, with these examples, rarefaction is by the tips of the tree and the resulting curve gives an indication of tree shape and distribution of sample observations amongst the tips of the tree.

One way to obtain summary statistics such as expectation and variance under rarefaction is to compute these statistics on samples drawn using a Monte Carlo procedure, that is, calculate the desired statistics on a collection of random draws.
On the other hand, there are closed form solutions for the mean of many measures of biodiversity under rarefaction.
For example, an analytical solution is well-known for species diversity, can be calculated for rarefaction by individuals and samples, and is much more efficient than resampling \citep{Hurlbert:1971tm,Ugland:2003tx,chiarucci2008discovering}.
However, we are not aware of such a formula for any phylogenetic diversity metrics.

In this paper, we establish analytical formulae for the mean and variance of phylogenetic diversity under rarefaction.
We develop these formulas in the setting of a phylogenetic tree with ``marks,'' which are a simple generalization allowing multiplicity of observations and arbitrary positions of observations along the tree.

\section{Materials and methods}

There are two different notions of the induced phylogenetic diversity (PD) of a subset $K$ of the leaves $L$ of a tree $T$; these notions have been called \emph{unrooted} and \emph{rooted} PD \citep{pardi2007resource}.
Unrooted PD is the total branch length of the smallest unrooted subtree contained in $T$ that has all of the leaves in $K$.
Rooted PD is the total branch length of the smallest rooted tree containing the original root of $T$ as well as the selected leaves $K$.
The rooted definition was that originally intended by \citet{Faith:1992vm}: see \citep{Faith:2006wp} for a historical discussion.
These two need not be the same: for example, any $K$ consisting of a single element will have zero unrooted phylogenetic diversity, but nonzero rooted phylogenetic diversity.
It is important to make a distinction between rooted and unrooted PD versus rooted and unrooted trees.
In our formulation, we are effectively treating all trees as rooted, by assigning an arbitrary root if necessary, and that unrooted and rooted PD refers specifically to the forced inclusion (or not) of a special root.

The two definitions of PD are useful in different domains of application.
For example, for conservation applications keeping a single species has significant value, thus it makes sense to have nonzero PD for a single species.
On the other hand when comparing the level of ecological diversity between environments, it may not make sense to keep the root, in which case the diversity between the members of a set of size one is zero.

We will derive formulae for both definitions of PD\@.
However, the description of the variance of unrooted PD will be deferred to the Appendix.

Formulae for rarefaction of phylogenetic diversity can be easily and productively generalized from the notion of a tree to the notion of a \emph{tree with marks}, which allows more flexibility in abundance weighting and attachment locations.
We define a \emph{tree with marks} as a tree along with a collection of special points on the tree (marked with stars in Fig.~\ref{FIGdefs}), which may be present with multiplicity.
The induced subtree of a collection of marks on a phylogenetic tree is the smallest connected set that contains all of those marks.
The phylogenetic diversity of a (sub)tree is the total branch length of the tree.

In this setting, marks represent observations.
Thus if a certain leaf taxon $t$ is observed $x$ times, $x$ marks are put at $t$.
However, it is just as easy to generalize to the setting where marks appear on the interior of tree edges.
The motivation for working in terms of marks is that it formalizes the notion of observation count and affords some extra flexibility for location of observations.
In particular, microbial ecologists often census a given community by high-throughput sequencing, and it is not practical to build a phylogenetic tree on all of the sequences thus created.
For this reason, sometimes scientists map sequences to trees using either similarity search plus a most recent common ancestor strategy, as in the work of \citet{huson2007megan}, or ``place'' the sequences into the tree using a phylogenetic criterion \citep{berger2011performance, matsen2010pplacer}.
The attachment point of a mapping of a sequence into a tree is then considered as a mark.

The \emph{unrooted phylogenetic diversity} of a tree with marks is the total branch length of the tree induced by those marks, that is, the total branch length of the smallest connected subset of the tree containing the marks.
The \emph{rooted phylogenetic diversity} of a rooted tree with marks is equal to the unrooted phylogenetic diversity of the tree with the given marks along with a mark added at the root; in this case the path from the root to the selected leaves is always included in the PD calculation.
These are simple generalizations of the corresponding definitions for leaf observations.

The following sections will be concerned with rarefying the collection of marks and computing phylogenetic diversities of the corresponding induced subtrees.
We will use \emph{proximal} to indicate the direction towards the root, and  \emph{distal} to mean the opposite.
If $T$ is unrooted, we will still use these terms for descriptive purposes; in this case an arbitrary root can be permanently assigned.

We fix a non-empty collection $M$ of $n$ marks on a tree $T$, and some number $1 \le k < n$ of marks to sample for our rarefaction.
Again, marks can be present multiple times in a collection, enabling the expression of multiplicity of observation of a taxon or sequence.
\begin{defn}
  Define an \emph{edge snip} to be a maximal segment of an edge with no marks or internal nodes.
\end{defn}
Say there are $s$ snips on the tree with marks, and that they are indexed by $i$.
Let $\ell_i$ be their length for $1 \leq i \leq s$.
Let $C_i$ be the set of marks that are proximal to snip $i$, and $D_i$ be the set that are distal to snip $i$ (Fig.~\ref{FIGdefs}).

\arxiv{\FIGdefs}

\begin{defn}
  For every $1 \leq i \leq s$, let $\Xr_i$ be the random variable that is equal to one if there is at least one mark on the distal side of snip $i$ after rarefaction, and zero otherwise.
  Let $\Xu_i$ be the random variable that is equal to one if there is at least one mark on each side of snip $i$ after rarefaction, and zero otherwise.
\end{defn}

The following two statements are true for $X \in \{\Xr, \Xu\}$ with the corresponding $Y \in \{\Yr, \Yu\}$.
The phylogenetic diversity $Y$ after rarefaction can be expressed as the random variable
\begin{equation}
  \label{eq:Yexp}
  Y = \sum_i \ell_i X_i.
\end{equation}
because the length of a snip $i$ contributes to the PD exactly when the corresponding $X_i = 1$.

Thus $\EE(Y) = \sum_i \ell_i \EE(X_i)$, and
\begin{equation}
  \label{eq:Yvar}
  \Var(Y) = \sum_{i,j} \ell_i \ell_j \Cov(X_i, X_j).
\end{equation}

To calculate expectations and covariances of the $X_i$, the following definition will be useful.
Fix an $R \subset M$.
Let $q_k(R)$ be the probability that nothing in $R$ is selected in a uniform sample of size $k$ from $M$ without replacement.
Recalling that $n = |M|$, note that (from the hypergeometric distribution):
\[
  q_k(R) =
    \begin{cases}
      \binom{n - |R|}{k} / \binom{n}{k} & \text{when $n - |R| \ge k$} \\
      0                                 & \text{otherwise}
    \end{cases}
\]
with the convention that $\binom{x}{0} = 1$ for all $x \in \mathbb{N}$.

Note that the $q_k(R)$ can be calculated for successive $k$ by observing that
\[
  q_{k+1}(R) = \frac{n-|R|-k}{n-k} \, q_k(R).
\]
Because the $q_k$ only depend on the size of $R$, a computer implementation only needs to calculate the $q_k(R)$ once for any $R$ of a given size; the $q_k(R)$ notation was chosen for convenience.

\subsection{Rooted phylogenetic diversity}

As described above, rooted phylogenetic diversity does PD calculation while always including the root.
By \eqref{eq:Yexp} and \eqref{eq:Yvar} all that is needed is the mean and the covariance matrix of the $\Xr_i$'s.
Note that $\Xr_i$ is zero unless at least one element of $D_i$ is sampled, in which case it is one.
Thus
\begin{equation}
  \EE[\Xr_i] = 1 - q_k(D_i).
  \label{eq:EXri}
\end{equation}

$\Xr_i \Xr_j$ is zero unless the rarefaction samples at least one element of both $D_i$ and $D_j$, in which case it is one.
The probability that one or both of these are empty under rarefaction is $q_k(D_i) + q_k(D_j) - q_k(D_i \cup D_j)$, thus
\[
  \EE[\Xr_i \Xr_j] = 1 - q_k(D_i) - q_k(D_j) + q_k(D_i \cup D_j).
\]
By \eqref{eq:EXri},
\[
  \EE[\Xr_i] \EE[\Xr_j] = 1 - q_k(D_i) - q_k(D_j) + q_k(D_i) q_k(D_j),
\]
thus,
\[
  \Cov(\Xr_i, \Xr_j) = q_k(D_i \cup D_j) - q_k(D_i) q_k(D_j).
\]

In summary,
\begin{align*}
  \EE[\Yr] & = \sum_i \ell_i \left[1 - q_k(D_i) \right]\\
  \Var[\Yr] & = \sum_{i,j} \ell_i \ell_j \left[ q_k(D_i \cup D_j) - q_k(D_i) q_k(D_j) \right].
\end{align*}

This solution can be seen to be a generalisation of the analytical formulae for the mean and variance of expected species richness under rarefaction \citep{Hurlbert:1971tm,HeckJr:1975tv} as follows.
Consider the special case of a ``star'' tree with all tips sharing a single common ancestor, where all marks are located at the tips of the tree (with the exception of one mark placed at the root), and where all branch lengths (and thus all snips) have a length of one.
Under these particular circumstances, the species richness and phylogenetic diversity of the collection of marks are equal and the formulae for mean and variance of expected phylogenetic diversity simplify to their equivalents for species richness.

\subsection{Unrooted phylogenetic diversity}

Assume as above that we are sampling $k>0$ marks for our rarefaction.
It is not possible for the rarefaction samples from both $C_i$ and $D_i$ to be empty.
Thus these two events are mutually exclusive, and
\begin{equation}
  \label{eq:EXui}
  \EE[\Xu_i] = 1 - q_k(C_i) - q_k(D_i).
\end{equation}

Then, by \eqref{eq:Yexp} and \eqref{eq:EXui},
  \begin{equation}
    \EE(\Yu) = \sum_{i} \ell_i \left[1 - q_k(C_i) - q_k(D_i) \right].
  \end{equation}

The variance of the unrooted case is deferred to the Appendix.

\subsection{Example applications}

We demonstrate our method for calculating the mean and variance of phylogenetic diversity under rarefaction using three examples.
In the first, we compare the rarefaction curve generated by Monte Carlo randomisation to that calculated by the exact analytical solution.
The data are counts of stems of all woody shrubs in forty plots in Toohey Forest, Queensland, Australia.
Within each plot, all plant stems over 0.3 m and under 3.0 m were counted; this figure was used as an index of abundance.
All shrubs were identified to species and a composite phylogeny was compiled from multiple published trees; see \citep{Nipperess:2010kp} for a more detailed description of the data.
Stem counts were summed across all plots to produce a single value per species before rarefaction by individual stems.
Of the total of 582 stems, rarefied values were calculated for every multiple of 10 stems from 10 to 580.
For the Monte Carlo procedure, mean and variance of phylogenetic diversity were calculated from 2,000 random subsamples of size $k$ from the pool of 582 stems.

Our second example demonstrates rarefaction of phylogenetic diversity by units of species.
Phylogenetic diversity of extant mammals was calculated for each terrestrial ecoregion on the Australian continental shelf (that is, Australia along with Tasmania, New Guinea and offshore islands).
Terrestrial ecoregions are biogeographic units representing distinct species assemblages \citep{Olson:2001wv}.
Species lists of mammals for each ecoregion were sourced from the WildFinder database maintained by the World Wildlife Fund (http://www.worldwildlife.org/science/wildfinder/).
Evolutionary relationships were sourced from a species-level supertree of the world's mammals \citep{BinindaEmonds:2007ul}.
Because of the strong correlation between species richness and phylogenetic diversity, rarefaction allows for the comparison of ecoregions with the effect of spatial variation in species richness removed.
To do this, the expected phylogenetic diversity for a subset of 25 mammal species was calculated for each ecoregion.
The value of 25 was chosen because it was the minimum species richness for this set of ecoregions.

Our third example comes from the human microbiome.
We reanalyze a pyrosequencing dataset describing bacterial communities from women with bacterial vaginosis \citep{srinivasan2012bacterial}.
Bacterial vaginosis (BV) has previously been shown to be associated with increased microbial community diversity \citep{fredricks2005molecular}.
For this study, swabs were taken from 242 women from the Public Health, Seattle and King County Sexually Transmitted Diseases Clinic between September 2006 and June 2010 of which 220 samples resulted in enough material to analyze  (data available as Sequence Read Archive submission SRA051298).
Vaginal fluid for each specimen was also evaluated according to Nugent score, which provides a diagnostic score for BV which ranges from 0 (BV-negative) to 10 (BV-positive) based on presence and absence of bacterial morphotypes as viewed under a microscope \citep{nugent1991reliability}.
Selection of reference sequences and sequence preprocessing were performed using the methods described by \citet{srinivasan2012bacterial}.
452,358 reads passed quality filtering, with a median of 1,779 reads per sample (range: 523--2,366).
For this application, we investigated the shape of the rarefaction curves with respect to resampling.

\arxiv{\FIGmean}
\arxiv{\FIGvariance}

\section{Results}

There was a very high degree of correspondence between the analytical and Monte Carlo methods for the expected value and variance of phylogenetic diversity of the Toohey Forest dataset under rarefaction (Fig.~\ref{FIGmean} and~\ref{FIGvariance}; corresponding results for unrooted PD are not shown).
In this application the Monte Carlo estimate of the PD variance does not converge quickly to the exact value, as can be seen from the deviations of the points (generated from 2,000 Monte Carlo samples) from the curve in Fig.~\ref{FIGvariance}.
Such slow convergence provides further motivation for an exact formula.

Correcting phylogenetic diversity for the number of species present made a substantial difference to the ranking of terrestrial ecoregions in terms of their diversity (Fig.~\ref{FIGmammalmap}).
With unrarefied phylogenetic diversity, the three highest ranked ecoregions (Southeastern Papuan rainforests, Southern New Guinea lowland rainforests, Central range montane rainforests) are found in New Guinea.
However, when variation in species richness is taken into account by rarefaction, two of the three highest ranked ecoregions (Australian Alps montane grasslands, Naracoorte woodlands) were in temperate Australia.
Thus the rarefied version demonstrates high phylogenetic diversity for this data set relative to the number of species present for those regions.

\arxiv{\FIGmammalmap}

\arxiv{\FIGrarefactNugent}

The rarefaction curves for the vaginal samples shows a connection between the Nugent score of the sample and the shape of the curve (Fig.~\ref{FIGrarefactNugent}).
The rarefaction curves for low Nugent score samples tend to start low and stay low.
The high Nugent score samples typically start higher than low Nugent score samples, and stay high.


\section{Discussion}

We have presented exact formulae for the mean and variance of rooted and unrooted phylogenetic diversity under rarefaction.
This solution gives results that are indistinguishable from those given by Monte Carlo randomisation.
The analytical method is preferred both because its results are exact and can be more efficient than sampling.

Rarefaction of phylogenetic diversity is seeing growing use in a variety of biological disciplines and we highlight two specific applications here.
Rarefaction of phylogenetic diversity by units of species allows for the assessment of phylogenetic diversity independent of species richness.
Removing the influence of species richness can allow for the fairer comparison of the evolutionary history of fauna and flora.
While it possible to make this correction by taking the residuals from a regression between species richness and PD \citep{forest2007preserving,davies2011phylogenetic}, the expected PD for a given species richness has also been determined by repeated subsampling of a species pool \citep{davies2006environmental,forest2007preserving,morlon2011spatial,Yu:2012jg}.
This latter method describes the relationship between phylogenetic diversity and species richness as a rarefaction curve.
Our example of the mammal faunas of the Australian continental shelf shows that such a correction can now be implemented with an exact analytical solution rather than repeated subsampling.
Further, as previously found by \citet{forest2007preserving} for the Cape flora of South Africa, correction for the number of species makes a substantial difference to the rank order of phylogenetic diversity of sites.

The rarefaction curves for the vaginal samples give interesting information about the distribution of phylotypes in the vaginal microbiome.
Some of this information recapitulates prior knowledge.
For example, samples with low Nugent score are typically dominated by a handful of bacterial species in the \emph{Lactobacillus} genus.
These rarefied curves start low and stay low.
If there are also other distantly-related organisms present, but in low abundance, the curve can start low and then curve up to a high level.
The high Nugent score samples, that tend to start high and increase rapidly, indicate that there are a considerable number of taxa spread across the tree that appear in the samples with nontrivial count.

Software implementing the exact analytical solution for rarefaction of phylogenetic diversity are already available. The \emph{phylorare} and \emph{phylocurve} functions are implemented in the R statistical environment \citep{R-Development-Core-Team:2012fk}.
These functions calculate mean rooted phylogenetic diversity and can be used to standardise a set of samples to a particular level of sampling effort (\emph{phylorare}) or generate a rarefaction curve for units of individuals, collections or species (\emph{phylocurve}).
These functions can be downloaded from http://davidnipperess.blogspot.com.au/.
The \pplacer\ suite of programs (http://matsen.fhcrc.org/pplacer/) is a collection of programs for ``phylogenetic placement'' and associated analyses.
The \guppy\ software is the main binary to perform downstream analysis of collections of placements.
It calculates Faith's phylogenetic diversity as well as a number of other phylogenetic diversity measures, including the abundance-weighted $\qbarD$ of \citet{chao2010phylogenetic}, a new one-parameter family of PD metrics (manuscript under preparation), phylogenetic entropy \citep{allen2009new}, and phylogenetic quadratic entropy \citep{rao1982diversity}.
It also calculates PD rarefaction curves with exact formulae as shown here, as well as those for phylogenetic quadratic entropy.

The work presented in this paper relates to and extends previous work in similar areas.
\citet{faller2008distribution} derived a central limit theorem for phylogenetic diversity under a model of random extinction.
In doing so, they also derived the mean and variance of phylogenetic diversity under this model.
This model is different than the setting of rarefaction in that the random variable signaling extinction is independent between species, which is not true for rarefaction to a given size considered here.

\citet{odwyer2012phylogenetic} have also independently calculated a mean and variance under sampling, but with a different focus: they consider the distributions that might be achieved through a variety of sampling schemes from the ``metacommunity tree'' of all extant lineages.
They derive the expressions for the mean and variance of phylogenetic diversity that we use as a starting point for our proofs and then apply them to their various sampling distributions, using an approximation to bound the variance above.
They consider the binomial, Poisson, and negative binomial distributions, but do not consider the hypergeometric distribution as done here, which corresponds to the case of sampling without replacement.
They do not derive exact expressions for the variance, nor do they consider unrooted PD or our more general setting.

Although we would like to extend the mean and variance formulas for PD under rarefaction to variants of PD, doing so may not be simple.
For example, it would be interesting to investigate the mean and variance of $\qbarD$, the abundance-weighted PD of \citet{chao2010phylogenetic}, under rarefaction.
For $q=0$, $\qbarD$ is PD divided by $T$; for $q=1$, $\qbarD$ is $\exp(H_p/T)$; for $q=2$, $\qbarD$ is $1/(1 - Q/T)$ where $T$ is the maximum height of the phylogenetic tree, $H_p$ is phylogenetic entropy \citep{allen2009new}, and $Q$ is quadratic entropy \citep{rao1982diversity,warwick1995new}.
Because these $q=1$ and $q=2$ cases are nonlinear functions of other abundance-weighted PD measures, the derivation of their mean and/or variance may be challenging.

Future work will include sensitivity of the PD rarefaction curve to tree shape and the distribution of individuals among species.
It would also be interesting to investigate extensions of the present work to the ``coverage-based'' framework recently proposed by \citet{chaocoverage}, as well as an extension to ``unconditional variance'' formulation of \citet{colwell2012models}.

\section{Acknowledgements}

DAN would like to thank Daniel Faith and Peter Wilson for discussions on probability and PD, Chris Longson and Peter Wilson for help with R programming, and Kyle Barton, Chris Burwell and Roger Kitching for the Toohey Forest dataset.
FAM gratefully acknowledges the programming assistance of Connor McCoy and Aaron Gallagher, and would like to thank David Fredricks, Noah Hoffman, Martin Morgan, and Sujatha Srinivasan at the Fred Hutchinson Cancer Research Center for sharing the bacterial vaginosis data.
Dan Faith, Anne Chao, and an anonymous reviewer provided helpful comments that greatly improved the manuscript.
DAN was funded by the Australian Research Council (DP0665761 and DP1095200), and FAM was funded by National Institutes of Health (R01 HG005966-01).

\newpage

\bibliography{rarefaction}
\bibliographystyle{alphaplainnat}

\newpage

\section{Appendix}

Here we will calculate the variance of the unrooted phylogenetic diversity $\Yu$ via \eqref{eq:Yvar}.
First note that $(\Xu_i)^2 = \Xu_i$ because $\Xu_i$ only takes the values 1 and 0.
Thus if $i=j$ then $\Cov(\Xu_i, \Xu_j) = \EE[\Xu_i] - \EE[\Xu_i]^2$ which can be calculated using \eqref{eq:EXui}.

Now assume $i \neq j$.
Instead of expressing the variance in terms of functions of $C_i$'s and $D_i$'s, we will express them in terms of the following quantities defined for a pair of edges $i$ and $j$.
If $i$ is proximal to $j$, let $\Sij$ be $D_i$ (which is the \underline{Same} side of $i$ as $j$), $\Oij$ be $C_i$ (which is on the \underline{Other} side of $i$ from $j$), $\Sji$ be $C_j$, and $\Oji$ be $D_j$.
If $j$ is proximal to $i$, the roles of $i$ and $j$ are reversed.
If the path from $i$ to $j$ traverses the root, then let $\Sij$ be $C_i$, $\Oij$ be $D_i$, $\Sji$ be $C_j$, and $\Oji$ be $D_j$.
Since $\Oij$ and $\Sij$ are just $C_i$ and $D_i$ in some order, we can think of $\Xu_i$ being the random variable equal to one if the rarefaction sample from both $\Oij$ and $\Sij$ are nonempty and zero otherwise; the corresponding definition is true for $\Xu_j$.
By these definitions, a key point is that $\Oij \subset \Sji$ and $\Oji \subset \Sij$.

To calculate $\Cov(\Xu_i, \Xu_j) = \EE[\Xu_i \Xu_j] - \EE[\Xu_i] \EE[\Xu_j]$, note that $\Xu_i \Xu_j$ is zero unless the rarefaction samples at least one element of both $\Oij$ and $\Oji$, in which case it is one.
The probability that one or both of these are empty under rarefaction is $q_k(\Oij) + q_k(\Oji) - q_k(\Oij \cup \Oji)$, thus (for $i \neq j$)
\[
  \EE[\Xu_i \Xu_j] = 1 - q_k(\Oij) - q_k(\Oji) + q_k(\Oij \cup \Oji).
\]
By \eqref{eq:EXui},
\begin{align*}
  \EE[\Xu_i] \EE[\Xu_j] & = [1 - q_k(\Oij) - q_k(\Sij)] [1 - q_k(\Oji) - q_k(\Sji)] \\
 & = 1 - q_k(\Oij) - q_k(\Oji) + q_k(\Oij) q_k(\Oji) \\
 & \qquad - [1 - q_k(\Oij)] q_k(\Sji) - q_k(\Sij) [1 - q_k(\Oji)] + q_k(\Sij) q_k(\Sji).
\end{align*}

Thus (again, for $i \neq j$),
\begin{align*}
  \Cov(\Xu_i, \Xu_j) = q_k(\Oij & \cup \Oji) - q_k(\Oij) q_k(\Oji) + [1 - q_k(\Oij)] q_k(\Sji) \\
    & + q_k(\Sij) [1 - q_k(\Oji)] - q_k(\Sij) q_k(\Sji).
\end{align*}

These expressions can then be substituted back into \eqref{eq:Yvar} to get an expression for the variance of phylogenetic diversity.

\notarxiv{\
\newpage
\FIGdefs
\newpage
\FIGmean
\newpage
\FIGvariance
\newpage
\FIGmammalmap
\newpage
\FIGrarefactNugent
\newpage
}

\end{document}